\documentclass[sigconf,natbib]{acmart}
\AtBeginDocument{
  \providecommand\BibTeX{{
    \normalfont B\kern-0.5em{\scshape i\kern-0.25em b}\kern-0.8em\TeX}}}

\copyrightyear{2022}
\acmYear{2022}
\setcopyright{acmcopyright}
\acmConference[CIKM '22] {Proceedings of the 31th ACM International Conference on Information and Knowledge Management}{October 17--21, 2022}{Atlanta, GA, USA}
\acmBooktitle{Proceedings of the 31th ACM International Conference on Information and Knowledge Management (CIKM '22), October 17--21, 2022, Atlanta, GA, USA}
\acmPrice{15.00}
\acmDOI{10.1145/3511808.3557607}
\acmISBN{978-1-4503-9236-5/22/10}

\AtBeginDocument{
  \providecommand\BibTeX{{
    \normalfont B\kern-0.5em{\scshape i\kern-0.25em b}\kern-0.8em\TeX}}}

\usepackage{caption}
\usepackage{multirow}

\begin{document}

\fancyhead{}

\title{GReS: Graphical Cross-domain Recommendation for Supply Chain Platform}

\author{Zhiwen Jing}
\email{jingzhiwen0426@link.tyut.edu.cn}
\orcid{0000-0003-0000-8937}
\authornotemark[1]
\affiliation{%
  \department{College of Information and Computer}
  \institution{Taiyuan University of Technology}
  \city{Taiyuan}
  \country{China}
}

\author{Ziliang	Zhao}
\email{zhaoziliang@ruc.edu.cn}
\affiliation{%
  \department{Gaoling School of Artificial Intelligence}
  \institution{Renmin University of China}
  \city{Beijing}
  \country{China}
}

\author{Yang Feng}
\email{fengyang12@meituan.com}
\affiliation{%
  \institution{Meituan}
  \city{Beijing}
  \country{China}
}

\author{Xaochen Ma}
\email{maxiaochen@meituan.com}
\affiliation{%
  \institution{Meituan}
  \city{Beijing}
  \country{China}
}

\author{Nan Wu}
\email{wunan25@meituan.com}
\affiliation{%
  \institution{Meituan}
  \city{Beijing}
  \country{China}
}

\author{Shengqiao Kang}
\email{kangshengqiao@meituan.com}
\affiliation{%
  \institution{Meituan}
  \city{Beijing}
  \country{China}
}

\author{Cheng Yang}
\email{yangyangyang3701@gmail.com}
\orcid{0000-0001-7507-6171}
\affiliation{%
  \department{School of Computer and Information Engineering}
  \institution{Zhejiang Gongshang University}
  \city{Hangzhou}
  \country{China}
}

\author{Yujia Zhang}
\email{zhangyujia0474@link.tyut.edu.cn}
\affiliation{%
  \department{College of Information and Computer}
  \institution{Taiyuan University of Technology}
  \city{Taiyuan}
  \country{China}
}

\author{Hao Guo}
\email{feiyu_guo@sina.com}
\affiliation{%
  \department{College of Information and Computer}
  \institution{Taiyuan University of Technology}
  \city{Taiyuan}
  \country{China}
}

\begin{abstract}

Supply Chain Platforms (SCPs) provide downstream industries with raw materials. Compared with traditional e-commerce platforms, data in SCPs is more sparse due to limited user interests. To tackle the data sparsity problem, one can apply Cross-Domain Recommendation (CDR) to improve the recommendation performance of the target domain with the source domain information. However, applying CDR to SCPs directly ignores hierarchical structures of commodities in SCPs, which reduce recommendation performance. In this paper, we take the catering platform as an example and propose GReS, a graphical CDR model. The model first constructs a tree-shaped graph to represent the hierarchy of different nodes of dishes and ingredients, and then applies our proposed Tree2vec method combining GCN and BERT models to embed the graph for recommendations. Experimental results show that GReS significantly outperforms state-of-the-art methods in CDR for SCPs.\footnote{Zhiwen Jing and Ziliang Zhao make equal contribution to this research.}.

\end{abstract}

\begin{CCSXML}
<ccs2012>
<concept>
<concept_id>10002951.10003317.10003347.10003350</concept_id>
<concept_desc>Information systems~Recommender systems</concept_desc>
<concept_significance>500</concept_significance>
</concept>
</ccs2012>
\end{CCSXML}

\ccsdesc[500]{Information systems~Recommender systems}

\keywords{Cross-Domain Recommendation, Supply Chain Platform}

\maketitle

\section{Introduction} \label{INTRO}

\begin{figure}[h]
\centering
\includegraphics[width=\linewidth]{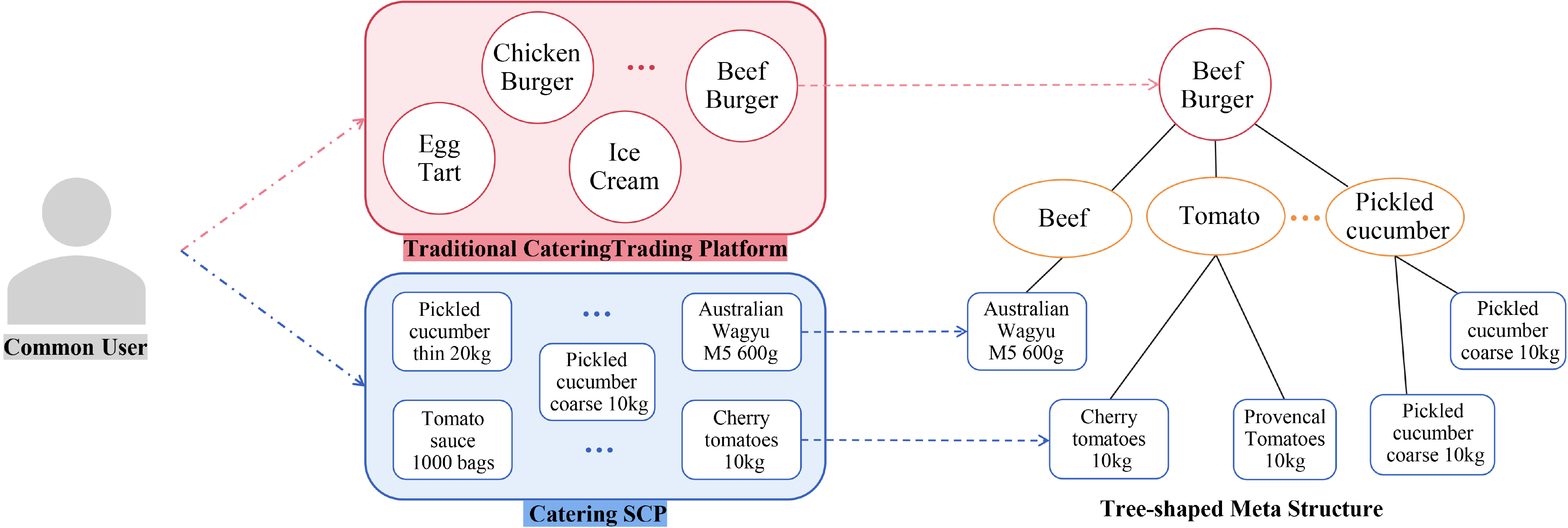}
\caption{CDR for Supply Chain Platform of Catering} \label{intro}
\end{figure}

The development of Supply-chain Platforms (SCPs) exacerbates the information overload of users, which emphasizes the importance of recommender systems for SCPs. However, due to users' concentrated interests and behaviors, recommendations in SCPs always go through the more severe data sparsity problem~\cite{sparsity} compared with traditional e-commerce. As one solution to tackle the data sparsity problem, Cross-Domain Recommendation (CDR)~\cite{CDR} has been widely studied. The core idea of CDR is leveraging a rich domain to improve the recommendation performance of a sparse domain~\cite{overview}. Based on this, many effective methods have been proposed, including content-based~\cite{content_base, tan2014_content_base, fernandez2014_content_base}, embedding-based~\cite{xue2017_emb_base, he2017_emb_base, emb_base, liu2020_emb_base}, and rating pattern-based approaches~\cite{rp_base, farseev2017_rp_base, zhu2020_rp_base}. Due to the high user overlap between SCPs and downstream industries, CDR is suitable for improving the recommendation performance of SCPs.

In this paper, we study the CDR for catering supply chain platforms. The methods and conclusions can be directly extended to other SCPs because their similar features. As shown in Figure~\ref{intro}, the CDR for catering usually contains a catering SCP (domain \emph{A}) and a traditional catering trading platform (domain \emph{B}). The common users first buy ingredients in domain \emph{A}, make these ingredients into dishes, and then sell these dishes in domain \emph{B}. In other words, the common users are consumers in domain \emph{A} and suppliers in domain \emph{B}. Therefore, it is reasonable to improve the recommendation performance in domain \emph{A} through the data in domain \emph{B}.

However, applying CDR to SCPs is challenging. First, there is an information gap between domain \emph{A} and \emph{B}. Taking ``Beef Burger'' as an example, the word segmentation and semantic understanding of ``Beef Burger'' in domain \emph{B} can only yield two original components, ``Beef'' and ``Burger'', but this is incomplete and inaccurate. In fact, the beef burger of domain \emph{B} consist of olive oil, onion, beef, etc. in domain \emph{A}, which is not well-considered. Second, there is a hierarchical relationship between dishes in domain \emph{B} and ingredients in domain \emph{A}. In traditional CDR~\cite{CTR-RBF, TMH, EMCDR, GA-DTCDR}, the relationships among products in different domains are equal. For example, if common users have watched many sci-fi movies in the movie domain, the model would recommend sci-fi novels to these users in the novel domain. However, in the CDR for SCPs, ``Pickled Cucumber'' and several other ingredients in domain \emph{A} constitute the ``Beef Burger'' in domain \emph{B} together, so the relationship between ``Pickled Cucumber'' and ``Beef Burger'' can be described hierarchically. The hierarchical relationship between dishes in domain \emph{B} and ingredients in domain \emph{A} is difficult to be represented properly.

To solve the challenges above, we propose GReS, a Graphical CDR model for SCPs. Specifically, to generate better embeddings for users and items in domain \emph{A}, we construct a Heterogeneous Graph (HG) with ratings and contents~\cite{GA-DTCDR}. We then apply Node2vec~\cite{Node2vec} to embed the HG for downstream recommendations. As for domain \emph{B}, we build several Tree-shaped Meta Structures (TMSs) of dishes, items (ingredients in domain \emph{A}), and item categories to describe the hierarchical relation among nodes. These TMSs constitute a {Tree-shaped Graph (TG)}. We then map the TG to embedding by a Tree2vec method that combines GCN model~\cite{GCN} and BERT model~\cite{BERT}. The GCN model focuses on extracting relational features in the tree-shaped graph, and the BERT model is suitable for extracting semantic features of the graph. We also use element-wise attention~\cite{GA-DTCDR} to effectively combine embeddings of users and items from two domains. Since there is no publicly available dataset in CDR for SCPs, we build a large dataset sampled from a commercial online system. The experimental results and ablation studies on the dataset show that GReS significantly outperforms the state-of-art baselines in various evaluation metrics. The effectiveness of GReS makes it suitable to be applied in CDR for SCPs in the future.

Overall, the contributions of this paper include: 
\begin{itemize}
\item To the best of our knowledge, GReS is the first CDR model designed to alleviate the data sparsity problem in SCPs.
\item To model the hierarchy between the commodity of downstream industry and supply chain, we design the TMS and propose a Tree2vec method to map the TMS to embeddings.
\item The experimental results on a commercial dataset demonstrate the efficacy of GReS over existing baselines.
\end{itemize}

\section{GReS Model}

\begin{figure}[tp]
\centering
\includegraphics[width=\linewidth]{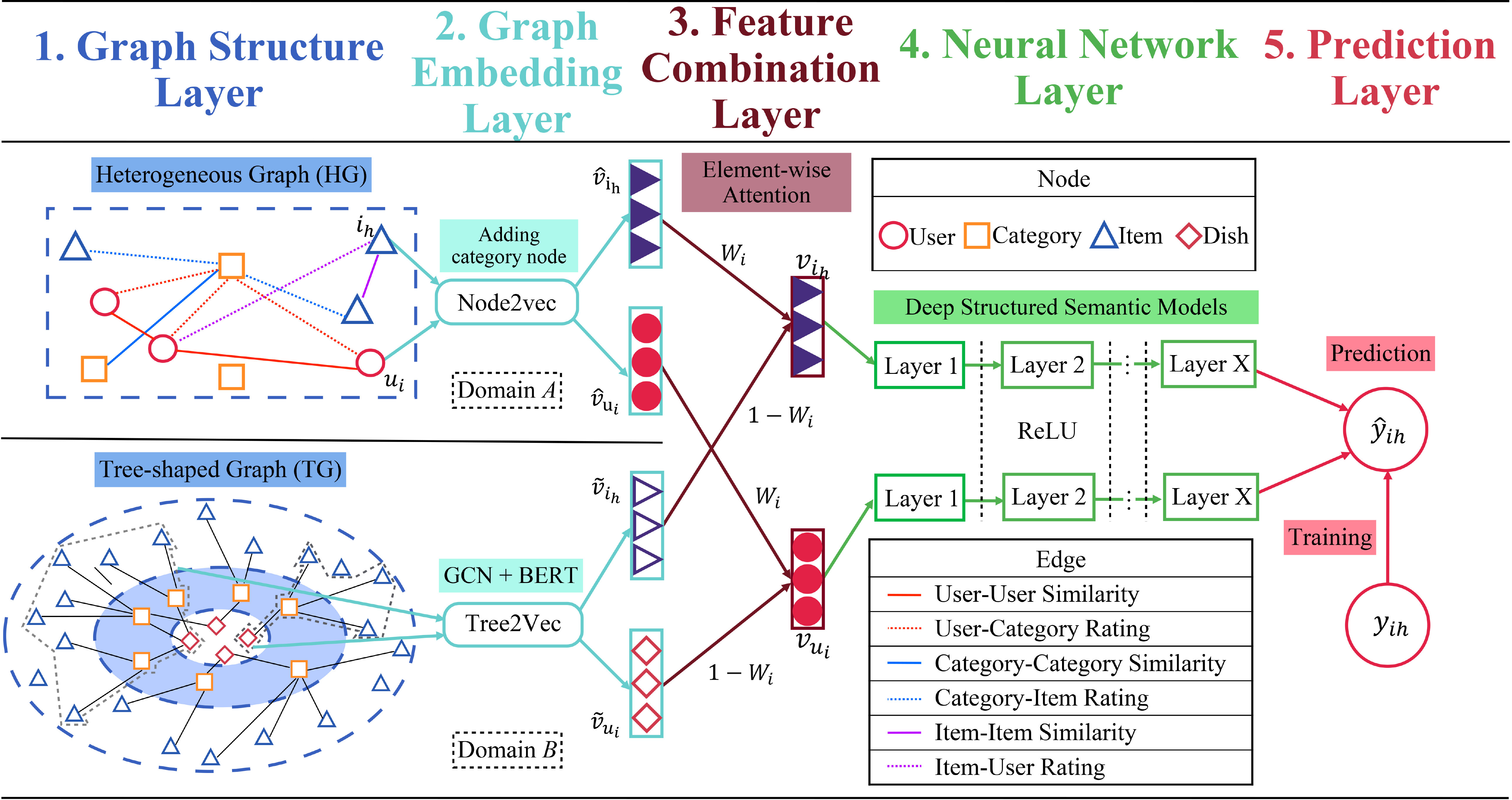}
\caption{The overall architecture of GReS model} \label{GRes}
\end{figure}

As illustrated in Figure~\ref{GRes}, GReS consists of five layers in total. First, in the graph structure layer, we use users and items to build two graphs, where domain \emph{A} is defined as a Heterogeneous Graph (HG), and domain \emph{B} is defined as a Tree-shaped Graph (TG). We then embed the two graphs using Node2vec~\cite{Node2vec} for domain \emph{A} and our proposed Tree2vec for domain \emph{B} in the graph embedding layer. Next, in the feature combination layer, we use the element-wise attention mechanism to concatenate embeddings from both domains. Finally, we apply an MLP to model the non-linear relationship between users and items and generate final user-item interaction predictions in the neural network layer and prediction layer.

\subsection{Graph Structure Layer} \label{gsl}

As shown in Figure~\ref{GRes}, the HG integrates information obtained from domain \emph{A} (SCPs). Based on existing work~\cite{GA-DTCDR}, we further introduce the item category node which is the category summary of items. For example, the item ``Australian Wagyu M5 600g'' belongs to the category ``Beef''. Introducing this node also adds three new edge types (user-category, category-category, and category-item) into existing types (user-user, item-item, and user-item). To construct the HG of domain \emph{A}, we apply the Doc2vec~\cite{Doc2vec} model to map the documents of user profiles, category details, and item details into vectors of users, categories, and items respectively. We then use the similarity between nodes to determine whether an edge among users, categories, and items should be generated. Taking users as an example (categories and items are built in a similar way), first, we calculate the Euclidean distance among users: $sim(i,j) = |Eu(u_i,u_j)| \label{sim}$,
where $u_i$ and $u_j$ are two different common users. Then we perform Min-Max normalization on $sim(i,j)$ to get $sim_M(i,j)$. After that, the edge weight between $u_i$ and $u_j$ can be calculated by:
\begin{equation}
    w(i,j) = \begin{cases}
  0, & sim_M(i,j) \le \alpha \\
  sim_M(i,j), & sim_M(i,j) > \alpha \\
\end{cases} \label{w}
\end{equation}
where $w(i,j)$ is the edge weight between $u_i$ and $u_j$, and $\alpha$ is a hyper-parameter which controls the sampling rate of the edge. For user-item, user-category, and category-item edges, we use the normalized order number of the items, categories, and items under each category as their corresponding edge weights.

To construct the TG of domain \emph{B}, it should consider common users who sell dishes in domain \emph{B} made by ingredients in domain \emph{A} (SCPs), including dishes, item categories, and items. Therefore, we design the TMS which is the component of TG to describe the hierarchical relationship of dishes-categories-items. Taking the dish``Beef Burger'' as an example, it is the parent of categories ``Beef'', ``Tomato'', and ``Pickled Cucumber'', and the category ``Beef'' is the parent of item ``Australian Wagyu M5 600g''. Therefore, we logically divide dishes, categories, and items of TG into three levels to form TMSs as shown in Figure~\ref{intro}. The weights in the TG will be determined in a similar way to the HG. % It is also worth noting that the TG is a directed graph that only allows information to be transferred from child nodes to parent nodes because a directed graph can emphasize the tree structure of TG.

\subsection{Graph Embedding Layer}

\begin{figure}[tp]
\centering
\includegraphics[width=\linewidth]{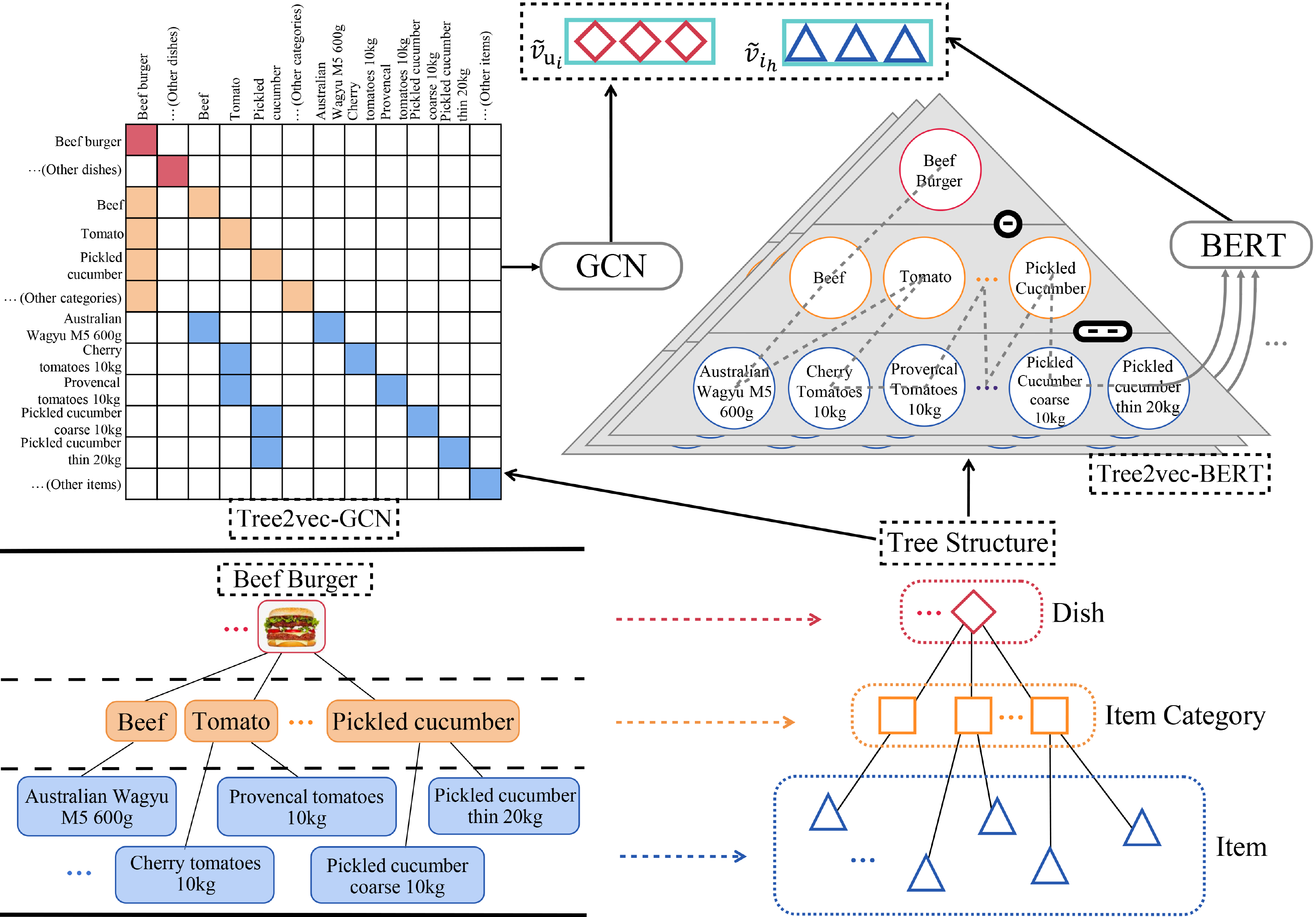}
\caption{Tree2vec model} \label{Tree2vec}
\end{figure}

Based on the HG of domain \emph{A}, we apply Node2vec~\cite{Node2vec} to generate an embedding for a user $u_i$ and an item $i_h$ as $\hat{v}_{u_i}$ and $\hat{v}_{i_h}$ respectively shown in Figure~\ref{GRes}. As for domain \emph{B}, it is crucial to leverage useful features in the TG for recommendation. In this paper, we mainly consider two types of features: spatial feature and semantic feature. The former concentrates on extracting spatial information to better encode the tree structure of the data. The latter focus on the interaction of semantics in the data. Given a common user $u_i$ and its TG $T_{u_i}$, the embedding of the TG is defined as:
$t_{i} = \text{GCN}(T_{u_i}) \oplus \text{BERT}(T_{u_i})$,
where ``$\oplus$'' is the concatenation operator. To implement the above two features, we propose a Tree2vec method containing a GCN model~\cite{GCN} for extracting spatial information and a BERT model~\cite{BERT} for extracting semantic information in the TMSs. The Tree2vec model then concatenates these two feature vectors as the embedding of the TG, and applies the embedding for recommendation. We ensure that a common user's corresponding TG has at least one TMS. That is, each merchant sells at least one dish.

\subsubsection{Tree2vec-GCN}

In Section~\ref{gsl}, we have aggregated all the TMSs of each common user and constructed an TG to represent the hierarchy of TMSs in domain \emph{B}. To extract spatial information, we apply GCN to encode the graph into a feature vector. Specifically, we first represent the TG as an adjacency matrix $M_T^i$. Each value in the matrix denotes the weight between different nodes. Since TG is a directed graph, the $M_T^i$ is a lower triangular matrix. After that, we implement a two-layers GCN model to encode $M_T^i$ as follows:

\begin{equation}
\begin{aligned}
H^{(l+1)}&=\sigma\left(\tilde{D}^{-\frac{1}{2}} \tilde{M_T^i} \tilde{D}^{-\frac{1}{2}} H^{(l)} W^{(l)}\right), \tilde{M_T^i} = M_T^i + I
\end{aligned}
\end{equation}
where $H^{(l)}$ is the matrix of node features in the $l$th layer of the GCN, $W$ is a trainable weight matrix, $\tilde{D}^{-\frac{1}{2}}$ is the degree matrix, and $\tilde{M_T^i}$ is the sum of $M_T^i$ and an identity matrix. We finally get the item embedding of the $h$th items $i_h$ as $G_{i_h}^{I}$, and add all dish embeddings of user $u_i$ to obtain the embedding of the $u_i$ as $G_{u_i}^{U}$.

\subsubsection{Tree2vec-BERT} 

The GCN model can only capture spatial structural information from the TG, while semantic information is not considered. Therefore, we further apply BERT language model~\cite{BERT} to learn semantic information of TMSs for each common user. To form a TMS of $u_i$ into a sequence, we apply the DFS strategy to traverse nodes in the entire TMS and form a node sequence $s_{u_i}^t$. To represent the hierarchical relationship of TMSs, during the traversal process, we add the following special tokens: (1) A ``-'' token denoting the edge from dish to category. (2) A ``- -'' token denoting the edge from category to item.

To distinguish different type of nodes, we further modify the position embeddings in original BERT model to denote three different levels, and use the pre-trained model BERT for representation learning of $s_{u_i}^t$ to get all dish embeddings and item embeddings. We then add all dish embeddings together to get the embedding of the common user $u_i$ as $B_{u_i}^{U}$. We directly use the item representation obtained from BERT as the item embedding of the $u_i$, denoted as $B_{i_h}^{I}$. Finally, we concatenate $G_{u_i}^{U}$ with $B_{u_i}^{U}$, and $G_{i_h}^{I}$ with $B_{i_h}^{I}$ respectively to obtain $u_i$'s embedding $\tilde{v} _{u_i}$ and $i_h$' embedding $\tilde{v} _{i_h}$.

\subsection{Feature Combination Layer}

In the feature combination layer, we use the element-wise attention mechanism to combine the user and item embeddings generated for two domains in the graph embedding layer. Taking the user as an example, for users with more sparse behaviour in domain \emph{A}, we combine the features with a larger proportion of the information in domain \emph{B}. User feature combination can be expressed as $v_{u_i} = W_i \cdot \hat{v} _{u_i} + (1 - W_i) \cdot \tilde{v} _{u_i} \label{v_ui}$, where $v_{u_i}$ is the combined feature vector of the user $u_i$, $W_i$ is a weight matrix in the attention network, and ``$\cdot$'' indicates element-wise product. The calculation process for the embedding of items is the same with that for users.

\section{Experiments and Analysis}

\subsection{Experiment}

We conduct experiments to answer the following key questions: \textbf{RQ1:} Does the TMS improve the recommendation performance? \textbf{RQ2:} How do GCN and BERT in Tree2vec contribute to improving recommendation performance? \textbf{RQ3:} What is the performance of GReS under different proportions of unique users?

\subsection{Experimental Settings}

\begin{table*}[] \small
\caption{Performance comparison in HR@$K$, NDCG@$K$ and MRR@$K$ where $K=5, 10, 20, 50$. The best scores and the second best scores are bold and underlined respectively. B-best and G-best are the best scores of baselines and variants of GReS.}
\begin{tabular}{@{}cl|cccc|cccc|cccc@{}}
\toprule
\multicolumn{2}{c|}{\multirow{2}{*}{Model}} & \multicolumn{4}{c|}{HR@$K$}          & \multicolumn{4}{c|}{NDCG@ $K$}       & \multicolumn{4}{c}{MRR@$K$}         \\ \cmidrule(l){3-14} 
\multicolumn{2}{c|}{}                       & $ K=5$ & $ K=10$ & $ K=20$ & $ K=50$ & $ K=5$ & $ K=10$ & $ K=20$ & $ K=50$ & $ K=5$ & $ K=10$ & $ K=20$ & $ K=50$ \\ \midrule
\multicolumn{2}{c|}{CTR-RBF}                & .0213  & .0334 & .0603 & .0811 & .0166 & .0234 & .0376 & .0462 & .0151 & .0211 & .0301 & .0350 \\
\multicolumn{2}{c|}{TMH}                    & .0224  & .0398 & .0687	& .0852 & .0179 &	.0278 & .0417 &	.0468 & .0169 &	.0251 & .0339 &	.0361 \\
\multicolumn{2}{c|}{EMCDR}                  & .0311  & .0667 & .0923 &	.1026 & .0246	& .0464 & .0545	& .0572 & .0226	& .0399 & .0447	& .0452 \\
\multicolumn{2}{c|}{GA-DTCDR}               & .0316  & .0659 & .1056 &	.1173 & .0250	& .0468 & .0597	& .0666 & .0224	& .0394 & .0478	& .0508 \\
\multicolumn{2}{c|}{DCDIR}                  & .0438  & .0953 & .1320 &	.1422 & .0334	& .0632 & .0794 &	.0829 & .0328	& .0533 & .0631	& .0633 \\
\multicolumn{2}{c|}{GReS}                   & \textbf{.0467}  & \underline{.1023} & \textbf{.1477} &	\textbf{.1577} & \textbf{.0360}	& \textbf{.0692} & \textbf{.0863}	& \textbf{.0889} & \textbf{.0344}	& \textbf{.0593} & \textbf{.0698}	& \textbf{.0707} \\ \hline
\multicolumn{2}{c|}{GReS w/o tree}          & .0353  & .0828 & .1303 &	.1388 & .0276 &	.0578 & .0747 & .0811 & .0253 &	.0482 & .0616	& .0623 \\
\multicolumn{2}{c|}{GReS w/o GCN}           & .0414  & .0878 & .1403	& .1487 & .0325	& .0601 & .0808 &	.0831 & .0313	& .0512 & .0631 &	.0641 \\
\multicolumn{2}{c|}{GReS w/o unidirectional}& .0442  & .0994 & .1423 & .1533 & .0351 &	.0659 & \underline{.0855} & .0872 & \underline{.0332} & .0553 & \underline{.0683} &	\underline{.0688} \\
\multicolumn{2}{c|}{GReS w/o BERT}          & .0404  & .0861 & .1421 &	.1455 & .0337 &	.0613 & .0821 &	.0838 & .0321 &	.0507 & .0627 & .0630 \\
\multicolumn{2}{c|}{GReS w/o fix-position}  & \underline{.0454}  & \textbf{.1031} & \underline{.1433} & \underline{.1551} & \underline{.0352} & \underline{.0690} & .0853 &	\underline{.0879} & .0331 & \underline{.0579} & .0680 &	.0686 \\ \midrule
\multicolumn{2}{c|}{GReS vs. B-best}          & 6.62\% & 7.35\%& 11.89\%&	10.90\%& 7.78\%&	9.49\%& 8.69\% &	7.24\% & 4.88\%&	11.26\%& 10.62\%&	11.69\%\\
\multicolumn{2}{c|}{GReS vs. G-best}        & 2.86\%  &-0.78\% & 3.07\% &	1.68\% & 2.27\% &	0.29\% & 0.94\%	& 1.14\% & 3.61\% &	2.42\% & 2.20\% &	2.76\% \\ \bottomrule
\end{tabular} 
\end{table*}

\subsubsection{Datasets}

There is no publicly available dataset in CDR for SCPs. To verify the effectiveness of GReS, we build a dataset based on a commercial daily log for experiments. The dataset consists of 3,000 common users, 600 unique users only belonging to the SCP, and their purchase and sales data in domain \emph{A} and domain \emph{B}.

\subsubsection{Comparative Baselines and Evaluation Metrics}

To prove the effectiveness of GReS, in our experiments, it is compared with five baselines which can be divided into three groups. They are Single-Domain Recommendation (SDR) (CTR-RBF~\cite{CTR-RBF}, TMH ~\cite{TMH}), CDR (EMCDR~\cite{EMCDR}, GA-DTCDR), and CDR with knowledge graph (DCDIR~\cite{DCDIR}). We also study some variants of GReS for Ablation Study. In order to be consistent with the previous work~\cite{CTR-RBF, TMH, EMCDR, GA-DTCDR, DCDIR}, all models are evaluated by HR@$K$, NDCG@$K$, and MRR@$K$.

\subsubsection{Implementation Details}

We randomly select 80\% data for training, 10\% for validation, and 10\% for testing. All parameters of the model are tuned by fitting the validation set and determined with an early-stop strategy. In the graph structure layer, we set the hyper-parameters of Doc2vec and Node2vec models as suggested in~\cite{Doc2vec, Node2vec}, and the sampling rate $\alpha$ as 0.05. For the performance of GReS under different proportions $M$ of unique users, the parameter $M \in $ \{0.010, 0.015, 0.020,0.025, 0.030, 0.035, 0.040, 0.045, 0.050\}. For three evaluation metrics HR@$K$, NDCG@$K$, and MRR@$K$, $K \in \{5, 10, 20, 50\}$. The neural network is trained using Adam~\cite{adam}, the learning rate is set to 0.0015, and the batch size is set to 30.

\subsection{Experimental Performance}

To verify the performance of GReS on SCPs (\textbf{RQ1}), we select five baselines for comparison. For Dual-Target CDR methods, we only discuss its performance on the sparsity domain to ensure the fairness of the comparison. In DCDIR~\cite{DCDIR}, we choose dish-category-item as the meta-path. Overall, GReS outperforms all baselines and is even 9.03\% averagely on three evaluation metrics better than the best DCDIR, indicating the effectiveness of our proposed Tree2vec model. The results also manifest that, (1) All CDR methods significantly outperform SDR methods, which proves that introducing relevant external features can alleviate the data sparsity problem of SDR. (2) The recommendation performance of DCDIR is generally better than other baselines with the benefit of representation of domain \emph{B} by knowledge graph. (3) With the increase of $K$, the speed of the improvement of all metrics slows down.

% \begin{figure}[h]
% \centering
% \includegraphics[width=\linewidth]{line.pdf}
% \caption{Performance comparison of different user proportions in $K=10$} \label{line}
% \Description{}
% \end{figure}

\subsection{Ablation Study}

To answer \textbf{RQ2}, we compare GReS with its five variants. We directly use the embedding of the TG without TMSs and denote it as GReS w/o tree. We removed GCN part and BERT part in Tree2vec as GReS w/o GCN and GReS w/o BERT respectively. GReS w/o unidirectional means allowing nodes in domain \emph{B} passing message bidirectionally in the GCN. GReS w/o fix-position means that the position embedding of BERT will not be modified. The results show that, first, GReS w/o tree is inferior to GReS with an average of 17.36\%, which proves that the TMSs have a positive effect on the recommendation performance of SCPs. Second, benefit from the extraction of structural information from each user’s directed graph, the average performance of GReS is 9.57\% higher than GReS w/o GCN. And because the unidirectional passing of message depicts the hierarchical relationship of different nodes to a certain extent, the average performance of GReS is 3.25\% better than GReS w/o unidirectional. Finally, since GReS w/o BERT is difficult to capture the semantic information, its average performance is inferior to GReS by 9.68\%, which highlights the importance of BERT. Furthermore, for GReS w/o fix-position, its average performance drops by 1.80\% compared to GReS. It can be seen that setting fixed position embeddings for nodes at different levels is more conducive to node representation learning.

% \subsection{The impact of unique user proportion}

% Since different supply chains have different proportions of unique users, which would have an influence on the recommendation performance, we answer \textbf{RQ3} by experimenting with different proportions of unique users, where the proportion parameter $M$ ranges from 0.01 to 0.05 with a step of 0.005. Performance trends of different $K$ are similar, so we only report HR@$10$, NDCG@$10$, and MRR@$10$ in the experiment. Figure~\ref{line} shows that, as the proportion of unique users increases, the recommendation performance of each model decreases. However, the performance of GReS is always better than other models, with an average performance of 8.51\%, 1.41\% higher than the best baseline and best variant.

\section{Conclusion}

To alleviate the data sparsity problem in supply chain platforms, in this paper, we take catering SCP as an example and propose the GReS model. Specifically, for the target domain, we use a heterogeneous graph with three kinds of nodes to model users' behaviour data. For the source domain, we represent the hierarchical relationship by a tree-shaped graph composed of tree-shaped meta structures, and use the element-attention mechanism to concatenate the features of the source and target domains. Furthermore, we designed experiments to verify the performance of GReS on SCPs. On the other hand, due to the high similarity between different SCPs, we also believe that GReS still has an excellent performance in other SCP recommendation environments, such as industrial accessories, medical care, and commodity wholesale.

\begin{acks}

The work is supported by National Natural Science Foundation of China (61876124, 61873178) and Shanxi Provincial Department of Science and Technology Basic Research Project (20210302123129).

\end{acks}

\newpage

\bibliographystyle{ACM-Reference-Format}
\bibliography{sample-base}

%%% -*-BibTeX-*-
%%% Do NOT edit. File created by BibTeX with style
%%% ACM-Reference-Format-Journals [18-Jan-2012].

\begin{thebibliography}{23}

%%% ====================================================================
%%% NOTE TO THE USER: you can override these defaults by providing
%%% customized versions of any of these macros before the \bibliography
%%% command.  Each of them MUST provide its own final punctuation,
%%% except for \shownote{}, \showDOI{}, and \showURL{}.  The latter two
%%% do not use final punctuation, in order to avoid confusing it with
%%% the Web address.
%%%
%%% To suppress output of a particular field, define its macro to expand
%%% to an empty string, or better, \unskip, like this:
%%%
%%% \newcommand{\showDOI}[1]{\unskip}   % LaTeX syntax
%%%
%%% \def \showDOI #1{\unskip}           % plain TeX syntax
%%%
%%% ====================================================================

\ifx \showCODEN    \undefined \def \showCODEN     #1{\unskip}     \fi
\ifx \showDOI      \undefined \def \showDOI       #1{#1}\fi
\ifx \showISBNx    \undefined \def \showISBNx     #1{\unskip}     \fi
\ifx \showISBNxiii \undefined \def \showISBNxiii  #1{\unskip}     \fi
\ifx \showISSN     \undefined \def \showISSN      #1{\unskip}     \fi
\ifx \showLCCN     \undefined \def \showLCCN      #1{\unskip}     \fi
\ifx \shownote     \undefined \def \shownote      #1{#1}          \fi
\ifx \showarticletitle \undefined \def \showarticletitle #1{#1}   \fi
\ifx \showURL      \undefined \def \showURL       {\relax}        \fi
% The following commands are used for tagged output and should be
% invisible to TeX
\providecommand\bibfield[2]{#2}
\providecommand\bibinfo[2]{#2}
\providecommand\natexlab[1]{#1}
\providecommand\showeprint[2][]{arXiv:#2}

\bibitem[\protect\citeauthoryear{Berkovsky, Kuflik, and Ricci}{Berkovsky
  et~al\mbox{.}}{2007}]%
        {CDR}
\bibfield{author}{\bibinfo{person}{Shlomo Berkovsky}, \bibinfo{person}{Tsvi
  Kuflik}, {and} \bibinfo{person}{Francesco Ricci}.}
  \bibinfo{year}{2007}\natexlab{}.
\newblock \showarticletitle{Cross-domain mediation in collaborative filtering}.
  In \bibinfo{booktitle}{\emph{International Conference on User Modeling}}.
  Springer, \bibinfo{pages}{355--359}.
\newblock


\bibitem[\protect\citeauthoryear{Bi, Song, Yao, Wu, Wang, and Xiao}{Bi
  et~al\mbox{.}}{2020}]%
        {DCDIR}
\bibfield{author}{\bibinfo{person}{Ye Bi}, \bibinfo{person}{Liqiang Song},
  \bibinfo{person}{Mengqiu Yao}, \bibinfo{person}{Zhenyu Wu},
  \bibinfo{person}{Jianming Wang}, {and} \bibinfo{person}{Jing Xiao}.}
  \bibinfo{year}{2020}\natexlab{}.
\newblock \showarticletitle{DCDIR: A deep cross-domain recommendation system
  for cold start users in insurance domain}. In
  \bibinfo{booktitle}{\emph{Proceedings of the 43rd international ACM SIGIR
  conference on research and development in information retrieval}}.
  \bibinfo{pages}{1661--1664}.
\newblock


\bibitem[\protect\citeauthoryear{Devlin, Chang, Lee, and Toutanova}{Devlin
  et~al\mbox{.}}{2018}]%
        {BERT}
\bibfield{author}{\bibinfo{person}{Jacob Devlin}, \bibinfo{person}{Ming-Wei
  Chang}, \bibinfo{person}{Kenton Lee}, {and} \bibinfo{person}{Kristina
  Toutanova}.} \bibinfo{year}{2018}\natexlab{}.
\newblock \showarticletitle{Bert: Pre-training of deep bidirectional
  transformers for language understanding}.
\newblock \bibinfo{journal}{\emph{arXiv preprint arXiv:1810.04805}}
  (\bibinfo{year}{2018}).
\newblock


\bibitem[\protect\citeauthoryear{Farseev, Samborskii, Filchenkov, and
  Chua}{Farseev et~al\mbox{.}}{2017}]%
        {farseev2017_rp_base}
\bibfield{author}{\bibinfo{person}{Aleksandr Farseev}, \bibinfo{person}{Ivan
  Samborskii}, \bibinfo{person}{Andrey Filchenkov}, {and}
  \bibinfo{person}{Tat-Seng Chua}.} \bibinfo{year}{2017}\natexlab{}.
\newblock \showarticletitle{Cross-domain recommendation via clustering on
  multi-layer graphs}. In \bibinfo{booktitle}{\emph{Proceedings of the 40th
  international ACM SIGIR conference on research and development in information
  retrieval}}. \bibinfo{pages}{195--204}.
\newblock


\bibitem[\protect\citeauthoryear{Fern{\'a}ndez-Tob{\'\i}as and
  Cantador}{Fern{\'a}ndez-Tob{\'\i}as and Cantador}{2014}]%
        {fernandez2014_content_base}
\bibfield{author}{\bibinfo{person}{Ignacio Fern{\'a}ndez-Tob{\'\i}as} {and}
  \bibinfo{person}{Iv{\'a}n Cantador}.} \bibinfo{year}{2014}\natexlab{}.
\newblock \showarticletitle{Exploiting Social Tags in Matrix Factorization
  Models for Cross-domain Collaborative Filtering.}. In
  \bibinfo{booktitle}{\emph{CBRecSys@ RecSys}}. Citeseer,
  \bibinfo{pages}{34--41}.
\newblock


\bibitem[\protect\citeauthoryear{Grover and Leskovec}{Grover and
  Leskovec}{2016}]%
        {Node2vec}
\bibfield{author}{\bibinfo{person}{Aditya Grover} {and} \bibinfo{person}{Jure
  Leskovec}.} \bibinfo{year}{2016}\natexlab{}.
\newblock \showarticletitle{node2vec: Scalable feature learning for networks}.
  In \bibinfo{booktitle}{\emph{Proceedings of the 22nd ACM SIGKDD international
  conference on Knowledge discovery and data mining}}.
  \bibinfo{pages}{855--864}.
\newblock


\bibitem[\protect\citeauthoryear{He, Liao, Zhang, Nie, Hu, and Chua}{He
  et~al\mbox{.}}{2017}]%
        {he2017_emb_base}
\bibfield{author}{\bibinfo{person}{Xiangnan He}, \bibinfo{person}{Lizi Liao},
  \bibinfo{person}{Hanwang Zhang}, \bibinfo{person}{Liqiang Nie},
  \bibinfo{person}{Xia Hu}, {and} \bibinfo{person}{Tat-Seng Chua}.}
  \bibinfo{year}{2017}\natexlab{}.
\newblock \showarticletitle{Neural collaborative filtering}. In
  \bibinfo{booktitle}{\emph{Proceedings of the 26th international conference on
  world wide web}}. \bibinfo{pages}{173--182}.
\newblock


\bibitem[\protect\citeauthoryear{Hu, Zhang, and Yang}{Hu et~al\mbox{.}}{2019}]%
        {TMH}
\bibfield{author}{\bibinfo{person}{Guangneng Hu}, \bibinfo{person}{Yu Zhang},
  {and} \bibinfo{person}{Qiang Yang}.} \bibinfo{year}{2019}\natexlab{}.
\newblock \showarticletitle{Transfer meets hybrid: A synthetic approach for
  cross-domain collaborative filtering with text}. In
  \bibinfo{booktitle}{\emph{The World Wide Web Conference}}.
  \bibinfo{pages}{2822--2829}.
\newblock


\bibitem[\protect\citeauthoryear{Kingma and Ba}{Kingma and Ba}{2014}]%
        {adam}
\bibfield{author}{\bibinfo{person}{Diederik~P Kingma} {and}
  \bibinfo{person}{Jimmy Ba}.} \bibinfo{year}{2014}\natexlab{}.
\newblock \showarticletitle{Adam: A method for stochastic optimization}.
\newblock \bibinfo{journal}{\emph{arXiv preprint arXiv:1412.6980}}
  (\bibinfo{year}{2014}).
\newblock


\bibitem[\protect\citeauthoryear{Kipf and Welling}{Kipf and Welling}{2016}]%
        {GCN}
\bibfield{author}{\bibinfo{person}{Thomas~N Kipf} {and} \bibinfo{person}{Max
  Welling}.} \bibinfo{year}{2016}\natexlab{}.
\newblock \showarticletitle{Semi-supervised classification with graph
  convolutional networks}.
\newblock \bibinfo{journal}{\emph{arXiv preprint arXiv:1609.02907}}
  (\bibinfo{year}{2016}).
\newblock


\bibitem[\protect\citeauthoryear{Le and Mikolov}{Le and Mikolov}{2014}]%
        {Doc2vec}
\bibfield{author}{\bibinfo{person}{Quoc Le} {and} \bibinfo{person}{Tomas
  Mikolov}.} \bibinfo{year}{2014}\natexlab{}.
\newblock \showarticletitle{Distributed representations of sentences and
  documents}. In \bibinfo{booktitle}{\emph{International conference on machine
  learning}}. PMLR, \bibinfo{pages}{1188--1196}.
\newblock


\bibitem[\protect\citeauthoryear{Liu, Zhao, Zhuang, Liu, Sheng, Xu, Zhou, and
  Xiong}{Liu et~al\mbox{.}}{2020b}]%
        {emb_base}
\bibfield{author}{\bibinfo{person}{Jian Liu}, \bibinfo{person}{Pengpeng Zhao},
  \bibinfo{person}{Fuzhen Zhuang}, \bibinfo{person}{Yanchi Liu},
  \bibinfo{person}{Victor~S Sheng}, \bibinfo{person}{Jiajie Xu},
  \bibinfo{person}{Xiaofang Zhou}, {and} \bibinfo{person}{Hui Xiong}.}
  \bibinfo{year}{2020}\natexlab{b}.
\newblock \showarticletitle{Exploiting aesthetic preference in deep cross
  networks for cross-domain recommendation}. In
  \bibinfo{booktitle}{\emph{Proceedings of The Web Conference 2020}}.
  \bibinfo{pages}{2768--2774}.
\newblock


\bibitem[\protect\citeauthoryear{Liu, Li, Li, and Pan}{Liu
  et~al\mbox{.}}{2020a}]%
        {liu2020_emb_base}
\bibfield{author}{\bibinfo{person}{Meng Liu}, \bibinfo{person}{Jianjun Li},
  \bibinfo{person}{Guohui Li}, {and} \bibinfo{person}{Peng Pan}.}
  \bibinfo{year}{2020}\natexlab{a}.
\newblock \showarticletitle{Cross domain recommendation via bi-directional
  transfer graph collaborative filtering networks}. In
  \bibinfo{booktitle}{\emph{Proceedings of the 29th ACM International
  Conference on Information \& Knowledge Management}}.
  \bibinfo{pages}{885--894}.
\newblock


\bibitem[\protect\citeauthoryear{Man, Shen, Jin, and Cheng}{Man
  et~al\mbox{.}}{2017}]%
        {EMCDR}
\bibfield{author}{\bibinfo{person}{Tong Man}, \bibinfo{person}{Huawei Shen},
  \bibinfo{person}{Xiaolong Jin}, {and} \bibinfo{person}{Xueqi Cheng}.}
  \bibinfo{year}{2017}\natexlab{}.
\newblock \showarticletitle{Cross-domain recommendation: An embedding and
  mapping approach.}. In \bibinfo{booktitle}{\emph{IJCAI}},
  Vol.~\bibinfo{volume}{17}. \bibinfo{pages}{2464--2470}.
\newblock


\bibitem[\protect\citeauthoryear{Shapira}{Shapira}{2015}]%
        {sparsity}
\bibfield{author}{\bibinfo{person}{Bracha Shapira}.}
  \bibinfo{year}{2015}\natexlab{}.
\newblock \bibinfo{booktitle}{\emph{Recommender systems handbook}}.
\newblock \bibinfo{publisher}{Springer-verlag New York Incorporated}.
\newblock


\bibitem[\protect\citeauthoryear{Tan, Bu, Qin, Chen, and Cai}{Tan
  et~al\mbox{.}}{2014}]%
        {tan2014_content_base}
\bibfield{author}{\bibinfo{person}{Shulong Tan}, \bibinfo{person}{Jiajun Bu},
  \bibinfo{person}{Xuzhen Qin}, \bibinfo{person}{Chun Chen}, {and}
  \bibinfo{person}{Deng Cai}.} \bibinfo{year}{2014}\natexlab{}.
\newblock \showarticletitle{Cross domain recommendation based on multi-type
  media fusion}.
\newblock \bibinfo{journal}{\emph{Neurocomputing}}  \bibinfo{volume}{127}
  (\bibinfo{year}{2014}), \bibinfo{pages}{124--134}.
\newblock


\bibitem[\protect\citeauthoryear{Winoto and Tang}{Winoto and Tang}{2008}]%
        {content_base}
\bibfield{author}{\bibinfo{person}{Pinata Winoto} {and}
  \bibinfo{person}{Tiffany Tang}.} \bibinfo{year}{2008}\natexlab{}.
\newblock \showarticletitle{If you like the devil wears prada the book, will
  you also enjoy the devil wears prada the movie? a study of cross-domain
  recommendations}.
\newblock \bibinfo{journal}{\emph{New Generation Computing}}
  \bibinfo{volume}{26}, \bibinfo{number}{3} (\bibinfo{year}{2008}),
  \bibinfo{pages}{209--225}.
\newblock


\bibitem[\protect\citeauthoryear{Xin, Liu, Lin, Huang, Wei, and Guo}{Xin
  et~al\mbox{.}}{2015}]%
        {CTR-RBF}
\bibfield{author}{\bibinfo{person}{Xin Xin}, \bibinfo{person}{Zhirun Liu},
  \bibinfo{person}{Chin-Yew Lin}, \bibinfo{person}{Heyan Huang},
  \bibinfo{person}{Xiaochi Wei}, {and} \bibinfo{person}{Ping Guo}.}
  \bibinfo{year}{2015}\natexlab{}.
\newblock \showarticletitle{Cross-domain collaborative filtering with review
  text}. In \bibinfo{booktitle}{\emph{Twenty-Fourth International Joint
  Conference on Artificial Intelligence}}.
\newblock


\bibitem[\protect\citeauthoryear{Xue, Dai, Zhang, Huang, and Chen}{Xue
  et~al\mbox{.}}{2017}]%
        {xue2017_emb_base}
\bibfield{author}{\bibinfo{person}{Hong-Jian Xue}, \bibinfo{person}{Xinyu Dai},
  \bibinfo{person}{Jianbing Zhang}, \bibinfo{person}{Shujian Huang}, {and}
  \bibinfo{person}{Jiajun Chen}.} \bibinfo{year}{2017}\natexlab{}.
\newblock \showarticletitle{Deep matrix factorization models for recommender
  systems.}. In \bibinfo{booktitle}{\emph{IJCAI}}, Vol.~\bibinfo{volume}{17}.
  Melbourne, Australia, \bibinfo{pages}{3203--3209}.
\newblock


\bibitem[\protect\citeauthoryear{Zhang, Jin, Li, Ding, and Yang}{Zhang
  et~al\mbox{.}}{2016}]%
        {rp_base}
\bibfield{author}{\bibinfo{person}{Zihan Zhang}, \bibinfo{person}{Xiaoming
  Jin}, \bibinfo{person}{Lianghao Li}, \bibinfo{person}{Guiguang Ding}, {and}
  \bibinfo{person}{Qiang Yang}.} \bibinfo{year}{2016}\natexlab{}.
\newblock \showarticletitle{Multi-domain active learning for recommendation}.
  In \bibinfo{booktitle}{\emph{Thirtieth AAAI Conference on Artificial
  Intelligence}}.
\newblock


\bibitem[\protect\citeauthoryear{Zhu, Wang, Chen, Liu, Orgun, and Wu}{Zhu
  et~al\mbox{.}}{2020b}]%
        {zhu2020_rp_base}
\bibfield{author}{\bibinfo{person}{Feng Zhu}, \bibinfo{person}{Yan Wang},
  \bibinfo{person}{Chaochao Chen}, \bibinfo{person}{Guanfeng Liu},
  \bibinfo{person}{Mehmet Orgun}, {and} \bibinfo{person}{Jia Wu}.}
  \bibinfo{year}{2020}\natexlab{b}.
\newblock \showarticletitle{A deep framework for cross-domain and cross-system
  recommendations}.
\newblock \bibinfo{journal}{\emph{arXiv preprint arXiv:2009.06215}}
  (\bibinfo{year}{2020}).
\newblock


\bibitem[\protect\citeauthoryear{Zhu, Wang, Chen, Liu, and Zheng}{Zhu
  et~al\mbox{.}}{2020a}]%
        {GA-DTCDR}
\bibfield{author}{\bibinfo{person}{Feng Zhu}, \bibinfo{person}{Yan Wang},
  \bibinfo{person}{Chaochao Chen}, \bibinfo{person}{Guanfeng Liu}, {and}
  \bibinfo{person}{Xiaolin Zheng}.} \bibinfo{year}{2020}\natexlab{a}.
\newblock \showarticletitle{A Graphical and Attentional Framework for
  Dual-Target Cross-Domain Recommendation.}. In
  \bibinfo{booktitle}{\emph{IJCAI}}. \bibinfo{pages}{3001--3008}.
\newblock


\bibitem[\protect\citeauthoryear{Zhu, Wang, Chen, Zhou, Li, and Liu}{Zhu
  et~al\mbox{.}}{2021}]%
        {overview}
\bibfield{author}{\bibinfo{person}{Feng Zhu}, \bibinfo{person}{Yan Wang},
  \bibinfo{person}{Chaochao Chen}, \bibinfo{person}{Jun Zhou},
  \bibinfo{person}{Longfei Li}, {and} \bibinfo{person}{Guanfeng Liu}.}
  \bibinfo{year}{2021}\natexlab{}.
\newblock \showarticletitle{Cross-domain recommendation: challenges, progress,
  and prospects}.
\newblock \bibinfo{journal}{\emph{arXiv preprint arXiv:2103.01696}}
  (\bibinfo{year}{2021}).
\newblock


\end{thebibliography}

\end{document}